\documentclass[superscriptaddress,twocolumn,secnumarabic,
nobibnotes,aps,prd,showpacs,nofootinbib]{revtex4-1}
\usepackage{graphicx}
\usepackage{epsf}
\usepackage{bm}
\usepackage{amsmath}
\usepackage{amsfonts}
\usepackage{amssymb}
\usepackage{epstopdf}
\usepackage{natbib}
\usepackage{color}
\usepackage{color}
\usepackage{enumitem}
\providecommand{\U}[1]{\protect\rule{.1in}{.1in}}
\newcommand{\be}{\begin{equation}}
\newcommand{\ee}{\end{equation}}

\newcommand{\mincir}{\raise
-3.truept\hbox{\rlap{\hbox{$\sim$}}\raise4.truept\hbox{$<$}\ }}
\newcommand{\magcir}{\raise
-3.truept\hbox{\rlap{\hbox{$\sim$}}\raise4.truept\hbox{$>$}\ }}

\begin{document}

\title{Constraints from High-Precision Measurements of the Cosmic Microwave Background: The Case of Disintegrating Dark Matter with $\bm{\Lambda}$ or Dynamical Dark Energy}

\author{Wenzhong Liu}
\email{lwz\_lwzlwz@163.com}
\affiliation{Department of Physics, Liaoning Normal University, Dalian, 116029, P. R. China}

\author{Luis A. Anchordoqui}
\email{luis.anchordoqui@gmail.com}
\affiliation{Department of Physics and Astronomy,  Lehman College, City University of  New York, NY 10468, USA}
\affiliation{Department of Physics, Graduate Center, City University of New York,  NY 10016, USA}
\affiliation{Department of Astrophysics, American Museum of Natural History, NY 10024, USA}

\author{Eleonora~Di~Valentino}
\email{eleonora.di-valentino@durham.ac.uk}
\affiliation{Institute for Particle Physics Phenomenology, Department of Physics, Durham University, Durham DH1 3LE, UK}

\author{Supriya Pan}
\email{supriya.maths@presiuniv.ac.in}
\affiliation{Department of Mathematics, Presidency University, 86/1 College Street, Kolkata 700073, India}

\author{Yabo Wu} 
\email{ybwu61@163.com}
\affiliation{Department of Physics, Liaoning Normal University, Dalian, 116029, P. R. China}

\author{Weiqiang Yang}
\email{d11102004@163.com}
\affiliation{Department of Physics, Liaoning Normal University, Dalian, 116029, P. R. China}

\date{\today}

\preprint{}
\begin{abstract}
\noindent In recent years discrepancies have emerged in measurements of the present-day rate of expansion of the universe $H_0$ and in estimates of the clustering of matter $S_8$. Using the most recent cosmological observations we reexamine a novel model proposed to address these tensions, in which cold dark matter disintegrates into dark radiation. The disintegration process is controlled by its rate $Q = \alpha \mathcal{H} \rho_{\rm ddm}$, where $\alpha$ is a (constant) dimensionless parameter quantifying the strength of the disintegration mechanism and $\mathcal{H}$ is the conformal Hubble rate in the spatially flat  Friedmann-Lema\^{i}tre-Robertson-Walker universe and $\rho_{\rm ddm}$ is the energy density of the disintegrating cold dark matter. We constrain this model with the latest 2018 Planck temperature and polarization data, showing that there is no evidence for $\alpha \neq 0$  and that it cannot solve the $H_0$ tension below $3\sigma$, clashing with the result obtained by analyzing the Planck 2015 temperature data. We also investigate two possible extensions of the model in which the dark energy equation-of-state parameter $w \neq -1$. In this case it is possible to combine Planck data with the SH0ES measurement, and we demonstrate that in both these models the $H_0$ tension is resolved at the $1\sigma$ level, but the condition $w \neq -1$ exacerbates the $S_8$ tension. We also demonstrate that the addition of intermediate-redshift data (from the Pantheon supernova type Ia dataset and baryon acoustic oscillations) weakens the effectiveness of all these models to address the $H_0$ and $S_8$ tensions. 
\end{abstract}
\maketitle
\section{Introduction} \label{sec:intro}

Over the last decade or so, the successful operation of WMAP~\cite{Hinshaw:2012aka} and Planck~\cite{Akrami:2018vks} satellites together with large-scale galaxy surveys and observations from the Hubble space telescope have provided an avalanche of data, leaving no doubt that a new era is beginning for ``precision cosmology''. Currently, the concordance model of cosmology is $\Lambda$CDM. Within this model the expansion of the universe today is dominated by the cosmological constant $\Lambda$ and cold dark matter (CDM). Even though $\Lambda$CDM has become established as a well tested model, various discrepancies have emerged, including the more than $4\sigma$ tension between the observed~\cite{Riess:2019cxk,Riess:2020fzl} and inferred~\cite{Aghanim:2018eyx} values of the Hubble constant $H_0 \equiv 100 h~{\rm km/s/Mpc}$~\cite{DiValentino:2020zio}, as well as the discrepancy between the cosmological and local determination of $S_8 \equiv \sigma_8 \sqrt{\Omega_m/0.3}$~\cite{DiValentino:2020vvd} that quantifies the rms density fluctuations when smoothed with a top-hat filter of radius $8h^{-1}/{\rm Mpc} (\equiv \sigma_8)$ as a function of the present day value of the non-relativistic matter density parameter $\Omega_m$~\cite{DiValentino:2020zio,DiValentino:2020vvd}. Assuming a flat $\Lambda$CDM model the best-fit to extract cosmological parameters by the Planck Collaboration leads to
$H_0 = 67.27 \pm 0.60~{\rm km/s/Mpc}$ at 68\% CL~\cite{Aghanim:2018eyx}, whereas the SH0ES
Collaboration finds a larger value $H_0 = 73.2\pm  1.3~{\rm  km/s/Mpc}$~\cite{Riess:2020fzl}. On the assumption of $\Lambda$CDM the Planck Collaboration reported $S_8 = 0.830 \pm 0.013$~\cite{Aghanim:2018eyx}, which is in $3\sigma$ tension with KiDS-1000 data ($S_8 =
0.766^{+0.020}_{-0.014}$)~\cite{Asgari:2020wuj} and $3.4\sigma$ tension with a combination of BOSS and KV450 data ($S_8 =
0.728 \pm 0.026$)~\cite{Troster:2019ean}.  Systematic effects do not seem to be responsible for these discrepancies~\cite{Verde:2019ivm,Riess:2020sih,DiValentino:2020vnx} and thus a plethora of new cosmological models have been proposed to accommodate the data~\cite{DiValentino:2021izs,Perivolaropoulos:2021jda}.

The above mentioned discrepancies have become a common test-ground to uncover properties of the dark sector. New cosmological models modifying the dark sector now may include a period of early dark energy~\cite{Poulin:2018cxd,Karwal:2016vyq,Sakstein:2019fmf,Agrawal:2019lmo,Niedermann:2019olb,Niedermann:2020dwg,Ye:2020btb,Ye:2021iwa}, phantom dark energy~\cite{Yang:2018qmz,DiValentino:2019dzu,DiValentino:2020naf,Yang:2021flj,DiValentino:2021zxy}, interacting dark energy~\cite{Kumar:2016zpg,DiValentino:2019ffd,Kumar:2019wfs,Lucca:2020zjb,Yang:2019uog,DiValentino:2017iww,Yang:2020uga,Yang:2018euj,Anchordoqui:2019amx,Pan:2020bur,Pan:2019gop,Yang:2019uzo,Pan:2019jqh,Anchordoqui:2020sqo,Allali:2021azp,Nunes:2021zzi}, emergent dark energy~\cite{Li:2019yem,Pan:2019hac,Li:2020ybr,Hernandez-Almada:2020uyr,Benaoum:2020qsi,Yang:2021egn}, scattering-induced disintegrating dark matter~\cite{Bjaelde:2012wi,Bringmann:2018jpr,Pandey:2019plg,Xiao:2019ccl} and decaying dark matter~\cite{Berezhiani:2015yta,Benisty:2019pxb,Vattis:2019efj,Desai:2019pvs,Anchordoqui:2020djl,Blinov:2020uvz,Nygaard:2020sow,Chen:2020iwm}. It would be engaging and at the same time intriguing if both the $H_0$ and $S_8$ discrepancies were to be resolved  simultaneously, but as yet model building of the dark sector on this front has not been done to a satisfactory degree~\cite{Schoneberg:2021qvd,Anchordoqui:2021gji}. 

In this paper we reexamine the idea that scattering-induced disintegration of dark matter into dark radiation (dr), with an interaction rate proportional to the Hubble parameter, could re-accelerate the expansion rate to accommodate the $H_0$ tension. The relativistic degrees of freedom are generally parametrized via $N_{\rm eff}$~\cite{Steigman:1977kc} and constrained by observations of the cosmic microwave background (CMB). In the Standard Model, we have $N_{\rm eff}^{\rm SM} = 3.046$~\cite{Mangano:2005cc,deSalas:2016ztq,Akita:2020szl,Froustey:2020mcq,Bennett:2020zkv}, and so the disintegrating dark matter (ddm) model  would produce $\Delta N_{\rm eff} \equiv N_{\rm eff} - N_{\rm eff}^{\rm SM} >0$. At the same time the ddm would reduce the amount of CDM relaxing the $S_8$ tension.  Indeed, in Ref.~\cite{Pandey:2019plg} a likelihood analysis was carried out considering only 2015 Planck TT CMB data  at high multipoles and the local SH0ES prior on $H_0$. This study leads to a 1\% upper bound of CDM disintegration decreasing the $S_8$ tension down to $0.3\sigma$ while simultaneously increasing the central value of $H_0$ by a factor of about $1.023$. However, the addition of intermediate-redshift data from supernova and baryon acoustic oscillations (BAO) weakens the effectiveness of the ddm model, because when these data sets are included the upper bound on ddm reduces to about 0.5\% (bringing the $S_8$ tension  to roughly $1.5\sigma$) and the increment in the mean value of $H_0$ is only a factor of $1.005$ (alleviating the tension at the $2.5\sigma$ level). Herein we combine the ddm idea with various dark energy sectors including the dynamical dark energy as a generalized candidate of the dark energy. To constrain the models we adopt the Planck 2018 observations, which feature the addition of the high multipole polarization data that break the correlations among some model parameters, exacerbating the $H_0$ tension beyond the $3\sigma$ level.

Before proceeding, we pause to note that since a fraction of dark matter disintegrates into dr per Hubble time, the effect is amplified near the onset of matter domination and therefore becomes constrained by CMB data. Now, the baseline Planck 2015 dataset only contains information on the temperature spectrum, featuring a large correlation between $N_{\rm eff}$ and $H_0$. This implies that by accommodating a $\Delta N_{\rm eff} >0$ it is possible to naturally increase $H_0$.  To improve the determination of $N_{\rm eff}$, in our study we consider
the latest Planck 2018 data sample which contains both temperature and polarization measurements, together with a new optical depth $\tau$ estimate which strongly correlates with this parameter, and shifts the $N_{\rm eff}$ best fit towards lower values. Thereby, although $N_{\rm eff}$ affects the temperature spectrum modifying the smoothing in the damping tail and increasing the early Integrated Sachs-Wolfe (eISW) effect~\cite{Bowen:2001in,Hou_2013}, the polarization is not affected by the eISW effect, breaking the degeneracy while yielding a very robust measurement of dr~\cite{DiValentino:2018zjj}.\footnote{The well-known eISW effect encodes the contribution to CMB anisotropies originating in time-varying gravitational potentials at early times, shortly after recombination, when the universe was not entirely matter-dominated~\cite{Sachs:1967er,Rees:1968zza}.} The dr measurement is so robust, that even adding additional data (BAO or Pantheon) the central value does not shift from $3.046$.

The paper has been organized as follows. In Sec.~\ref{sec-2} we present the background and perturbations equations of the ddm scenarios. After that in Sec.~\ref{sec-data} we describe the observational datasets and the statistical methodology to constrain all the cosmological scenarios described in this article. Then, in Sec.~\ref{sec-results} we discuss the observational constraints extracted out of all the scenarios considered. Finally, in Sec.~\ref{sec-concl} we close this article with a brief summary of the results.

\section{Disintegrating Dark matter: Background and Perturbations}
\label{sec-2}

We consider a spatially flat Friedmann-Lema\^{i}tre-Robertson-Walker (FLRW) line element which provides a good description of our homogeneous and isotropic universe: 
\begin{equation}
ds^2 = -dt^2 + a^2(t) \, \delta_{ij} \, dx^i dx^j \,, 
\end{equation}
where $a (t)$ is the expansion scale factor of the universe. We expand the metric tensor into spatial average and small linear perturbations. We work in the synchronous gauge in conformal time $d\eta = dt/a$, for  which the scalar component of the inhomogeneous line element reduces to,
\begin{equation}
ds^2 = a^2(\eta) [-d\eta^2 + (\delta_{ij} + h_{ij}) \ dx^i dx^j] \,,
\end{equation}
with $h_{ij}$ a rank-2 symmetric tensor field associated to the six components of the perturbations of the spatial part. Following~\cite{Ma:1995ey}, $h_{ij}$ is parametrized in Fourier space $k$. We assume that the gravitational sector of the universe is perfectly described by the General Relativity and within the matter sector not all the components are independently conserved. Here we explore the cosmology of the disintegration mechanism between CDM and dr but the remaining components, namely baryons and dark energy are independently conserved.  
In the scattering-induced ddm model, the CDM (with equation of state parameter $w_{\rm ddm}=0$) disintegrates into dr (with $w_{\rm dr}=1/3$), and hence the continuity equations read:
\begin{eqnarray}
\rho'_{\rm ddm}+3\mathcal{H}\rho_{\rm ddm} &=& - Q, \label{cons:ddm1}\\
\rho'_{\rm dr}+4\mathcal{H}\rho_{\rm dr} &=& Q \label{cons:ddm2}\, ,
\end{eqnarray}
where $\rho_{\rm ddm}$ and $\rho_{\rm dr}$ are respectively the densities of  ddm and dr, the prime denotes the derivative with respect to the conformal time $\eta$, $Q = \alpha \mathcal{H} \rho_{\rm ddm}$ is the interaction rate that characterizes the disintegration process, $\mathcal{H}$ is the conformal Hubble rate, and $\alpha >0$ is the coupling parameter that quantifies the strength of the CDM disintegration into dark radiation.
Before proceeding, we pause to note that the assumption of $Q  \propto {\cal H} \rho_{\rm ddm}$ has {\it a priori} no physical basis\footnote{At this point the readers might be interested to know that in the interacting dark matter $-$ dark energy scenarios, some attempts have been made to justify the choice of the interaction functions, see for instance Refs.  \cite{Boehmer:2015kta,Boehmer:2015sha,DAmico:2016jbm,Kase:2019veo,Pan:2020zza,Pan:2020mst}.} and from the phenomenological point of view, the most general form of this interaction function may look like either $Q \equiv \mathcal{H} Q (\rho_{\rm ddm}, \rho_{\rm dr})$ or $Q  \equiv Q (\rho_{\rm ddm}, \rho_{\rm dr})$ (without the presence of the conformal Hubble factor), where $Q (\rho_{\rm ddm}, \rho_{\rm dr})$ is any abribrary function of $\rho_{\rm ddm}$, and $\rho_{\rm dr}$.
However, in this work we keep the choice of the interaction function as adopted in Refs.~\cite{Bjaelde:2012wi,Pandey:2019plg}
because we are interested in studying the impact of including the high-$\ell$ polarization Planck data into the analysis.

The evolution of the ddm and dr components at the background level  can be completely solved from Eqs.~(\ref{cons:ddm1}) and (\ref{cons:ddm2}), leading to:  
\begin{eqnarray}
\rho_{\rm ddm}&=& \rho_{{\rm ddm},0} \ a^{-(3+\alpha)}, \label{evol-cdm}\\
\rho_{\rm dr} &=& = \beta \ a^{-4} + \frac{\alpha}{1-\alpha}\rho_{\rm ddm,0} \ a^{-(3+\alpha)} \label{evol-dr}\, ,
\end{eqnarray}
where $\beta$ is a constant. For simplicity, herein we assume that the initial abundance of CDM $\equiv$ ddm is established at some early time $t_{\rm prod} \ll t_{\rm LS}$ and that $\Delta N_{\rm eff} \ll 1$ at $t_{\rm prod}$, where $t_{\rm LS}$ denotes the time of last scattering. All through the paper we adopt the subscript $0$ to indicate the 
quantities are evaluated today, with $a_0 =1$. Note that the first term in the right-hand-side of Eq.~(\ref{evol-dr}) behaves like a standard radiation density whereas the second term behaves like a fluid with an equation of state $\alpha/3$. Following~\cite{Bjaelde:2012wi,Pandey:2019plg}, we consider the case of a weak coupling between dark matter and dark radiation, i.e., $\alpha \ll 1$ which under the assumption that dark radiation is only produced via scattering-induced dark matter disintegration leads to $\beta \sim 0$.
With this in mind, the fraction of dark matter that disintegrates into dark radiation,
\begin{eqnarray}
f_{\rm ddm}=\frac{\rho_{{\rm dr},0}}{\rho_{{\rm ddm},0}}=\frac{\alpha}{1-\alpha},
\end{eqnarray}
remains constant over the time and the system of equations describing the evolution of CDM and dr are simplified to:
\begin{eqnarray}
\rho_{\rm ddm}&=& \rho_{\rm ddm,0} \ a^{-(3+\alpha)}, \label{ddm-new}\\
\rho_{\rm dr} &=& \frac{\alpha}{1-\alpha}\rho_{\rm ddm,0} \ a^{-(3+\alpha)} \label{dr-new}\, .
\end{eqnarray}
Now, for the total energy density of the universe given by $\rho_{\rm total}=\rho_{b}+\rho_{\rm ddm}+\rho_{\rm dr}+\rho_{\rm de}$, we can write down the evolution of the Hubble expansion as follows
\begin{widetext}
\begin{eqnarray}
\frac{H^2}{H_0^2} = \Omega_{b} \ a^{-3}+ \Omega_{\rm ddm} \ a^{-(3+\alpha)}+\frac{\alpha}{1-\alpha} \ \Omega_{\rm ddm}\ a^{-(3+\alpha)}
+\Omega_{\rm de} \ \exp\left( 3\int^1_a \frac{1+w_{\rm de}(a')}{a'}da' \right),
\end{eqnarray}
\end{widetext}
where $\Omega_i=\rho_{i,0}/\rho_{\rm crit,0}$ denotes the present-day density parameters, with $\rho_{\rm crit,0} = 3H_0^2/(8 \pi G)$ the  present-day  value  of  the  critical  density, and where $w_{\rm de} (a)$ is the barotropic equation of state of the dark energy, which could be either time independent or time dependent.\footnote{Because we are always referring to the present day density  parameters, we omit the subscript $0$ in this case.}  In this work we consider the two possibilities, with $w_{\rm de} (a)$ defined as follows: 
\begin{itemize}[noitemsep,topsep=0pt]
    \item Non-dynamical dark energy: We consider the cosmological constant $\Lambda$ (i.e. $w_{\rm de} (a) = -1$) as the canonical example of non-dynamical dark energy. The Hubble expansion in this case becomes, 
    \begin{widetext}
    \begin{eqnarray}
\frac{H^2}{H_0^2} = \Omega_{b} \ a^{-3}+\Omega_{\rm ddm} \ a^{-(3+\alpha)}+\frac{\alpha}{1-\alpha} \Omega_{\rm ddm} \ a^{-(3+\alpha)}
+\Omega_{\Lambda}\;, 
\end{eqnarray}
\end{widetext}
and we label this cosmological scenario as $\Lambda$ddm. 
\item Dynamical dark energy: In this category we assume two different dark energy candidates as follows
    \begin{itemize}[noitemsep,topsep=0pt]
        \item We consider the simplest dynamical dark energy model characterized by the constant equation of state $w_{\rm de} (a) = w_0 \neq -1$. The Hubble expansion in this case takes the form
        \begin{widetext}
      \begin{eqnarray}
\frac{H^2}{H_0^2} = \Omega_{b} \ a^{-3}+\Omega_{\rm ddm} \ a^{-(3+\alpha)}+\frac{\alpha}{1-\alpha}\Omega_{\rm ddm} \ a^{-(3+\alpha)}
+\Omega_{\rm de} \ a^{-3 (1+w_0)}\;, 
\end{eqnarray}
\end{widetext}
and we label this scenario as $w_0$ddm. 
\item As a general dynamical dark energy model,  we assume the most well known dynamical equation of state parametrization, namely the Chevallier-Polarski-Linder (CPL) parametrization
$w_{\rm de}(a)=w_0+w_a(1-a)$, 
where $w_0$ and $w_a = d w_{\rm de}(a)/da$ are the free parameters. The Hubble expansion here becomes, 
\begin{widetext}
\begin{eqnarray}
\frac{H^2}{H_0^2} = \Omega_{b} \ a^{-3}+\Omega_{\rm ddm} \ a^{-(3+\alpha)}+\frac{\alpha}{1-\alpha}\Omega_{\rm ddm} \ a^{-(3+\alpha)}
+\Omega_{\rm de} \ a^{-3(1+w_0+w_a)}\exp[-3w_a(1-a)]\;,
\end{eqnarray}
\end{widetext}
and we label this scenario as $w_0w_a$ddm. 
\end{itemize}
\end{itemize}

All in all, for the ddm scenario described by Eqs.~(\ref{cons:ddm1}) and (\ref{cons:ddm2}), we can explicitly solve its evolution at the background level for three specific dark energy equation of state parameters. Now, to address the impact of the ddm cosmological models on CMB and large-scale-structure (LSS) observables, we must not only account for the modified evolution of the background densities, but also include the effect of perturbations. Since dark energy does not interfere with the dark matter disintegration mechanism, the density and velocity perturbations for the dark energy fluid will be exactly the same as we have seen in the non-interacting cosmological models where dark energy is a component. For the perturbations in the cosmological fluid, the dimensionless density contrast $\delta_i = \delta \rho_i/\bar \rho_i$ conveniently describes the fluctuations in the energy density field of a given cosmological species $i$ and 
 $\theta_i$  the velocity divergence of the fluid with respect to the expansion, where the bar denotes background quantities. This means that the density perturbations and the velocity perturbations for the dark energy fluid assuming the synchronous gauge will respectively be given by the following set of equations~\cite{Ma:1995ey}: 
 \begin{widetext}
\begin{eqnarray}
\delta'_{\rm de} & = & - (1+w_{\rm de})\left(\theta_{\rm de}+ \frac{h'}{2}\right) - 3\mathcal{H}w'_{\rm de}\frac{\theta_{\rm de}}{k^2}
- 3 \mathcal{H} \left(c^2_s - w_{\rm de} \right) \left[ \delta_{\rm de}+3\mathcal{H}(1+w_{\rm de})\frac{\theta_{\rm de}}{k^2} \right], \\
\theta'_{\rm de} & = & - \mathcal{H}(1-3c^2_s)\theta_{\rm de} + \frac{c^2_s}{1+w_{\rm de}}k^2\delta_{\rm de} - k^2\sigma_{\rm de}
\end{eqnarray}
\end{widetext}
where $h \equiv h_{ii}$ denotes the trace part of the metric perturbation and $c^2_s$ is the physical sound speed of the dark energy in the rest frame. We have taken the usual assumption in which $c^2_s = 1$. Under this assumption we are considering that the dark energy does not cluster in the sub-Hubble scale. However, one can also consider $c_s^2$ to be a free parameter, but this parameter has been found to be unconstrained, see for instance~\cite{Bean:2003fb,Hannestad:2005ak,Corasaniti:2005pq,Xia:2007km}. Thus, the assumption of free $c_s^2$ does not offer any interesting physics. In what follows the shear perturbation of dark energy is taken to be $\sigma_{\rm de}$ = 0. The most vital changes that appear in the perturbations equations are due to the disintegration mechanism between CDM and dr.
 
Putting all this together the density and the velocity perturbation equations for the ddm and dr sectors are found to be
\begin{equation}
\delta'_{\rm ddm}  =  \theta_{\rm ddm}-\frac{h'}{2} \,, \label{eqnar1}
\end{equation}
\begin{equation}
\theta'_{\rm ddm}  =  -\mathcal{H} \ \theta_{\rm ddm} \,, \label{eqnar2} 
\end{equation}
\begin{equation}
\delta'_{\rm dr}  =  -\frac{4}{3}\left(\theta_{\rm dr}+\frac{h'}{2}\right)
+\alpha\mathcal{H}\frac{\rho_{\rm ddm}}{\rho_{\rm dr}}(\delta_{\rm ddm}-\delta_{\rm dr}) \,,
\end{equation}
\begin{equation}
\theta'_{\rm dr}  =  \frac{k^2}{4}\delta_{\rm dr}-\alpha\mathcal{H}\frac{3\rho_{\rm ddm}}{4\rho_{\rm dr}}\left(\frac{4}{3}\theta_{\rm dr}-\theta_{\rm ddm}\right)   -  k^2\sigma_{\rm dr} \, .
\label{eqnar3}
\end{equation}
Note that one can set $\theta_{\rm ddm}=0$ in the ddm comoving frame and with this consideration (\ref{eqnar1}) and (\ref{eqnar2}) simplify to: $\delta'_{\rm ddm} = - h'/2$ and $\theta'_{\rm ddm} = 0$. Hereafter, the shear perturbation of dark radiation is also taken to be $\sigma_{\rm dr} =0$.

\begin{figure}
	\centering
	\includegraphics[width=0.4\textwidth]{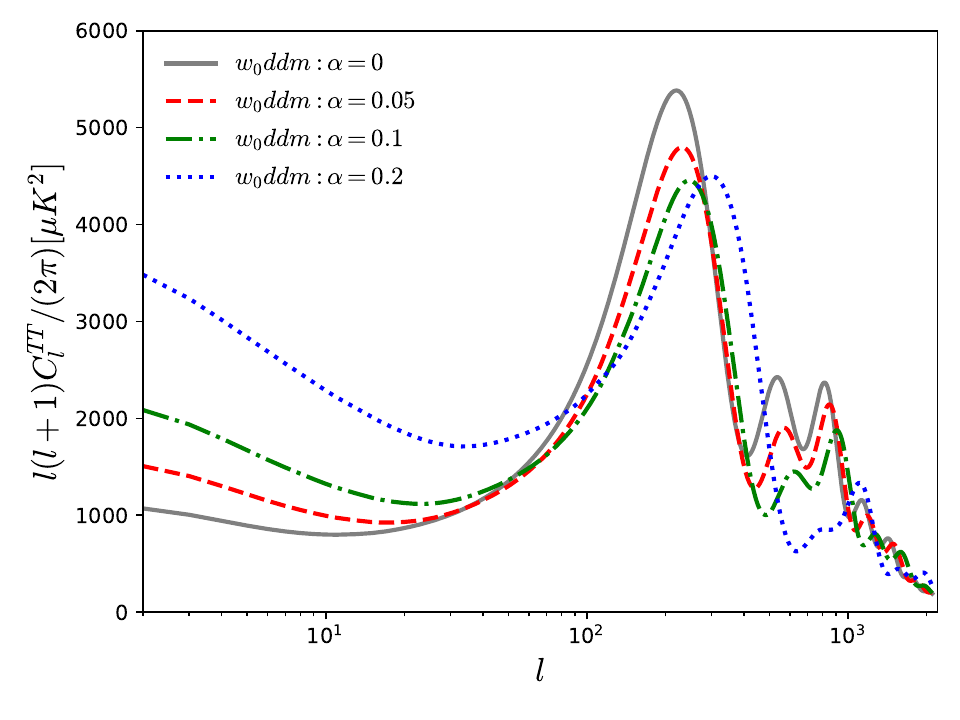}
	\includegraphics[width=0.4\textwidth]{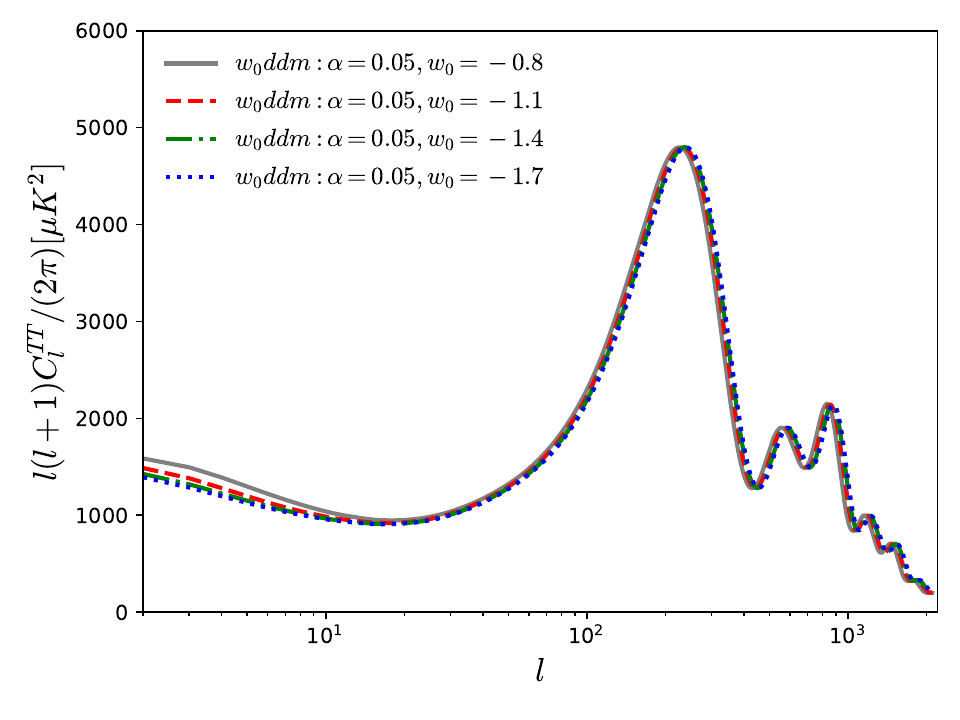}
	\caption{\textit{CMB TT power spectrum obtained by varying $\alpha$ (upper panel) and by varying $w_0$ (bottom panel) for the $w_0$ddm scenario.} }
	\label{fig:TT-w0ddm}
\end{figure}

\begin{figure}
	\centering
	\includegraphics[width=0.4\textwidth]{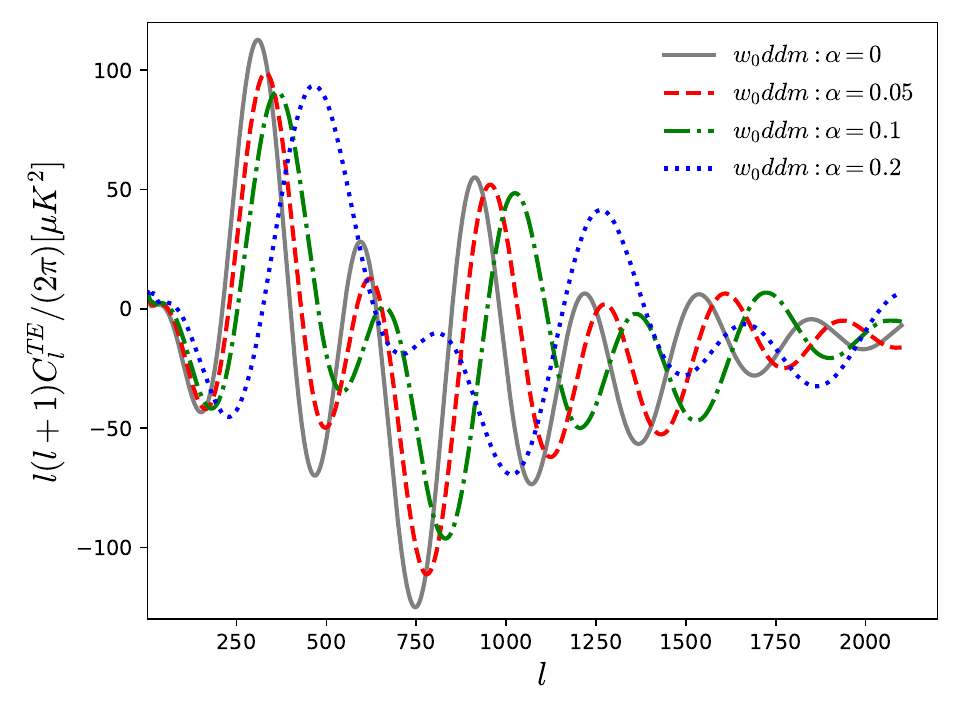}
	\caption{\textit{CMB TE power spectrum obtained by varying $\alpha$  for the $w_0$ddm scenario.} }
	\label{fig:TE-w0ddm}
\end{figure}

\begin{figure}
	\centering
	\includegraphics[width=0.4\textwidth]{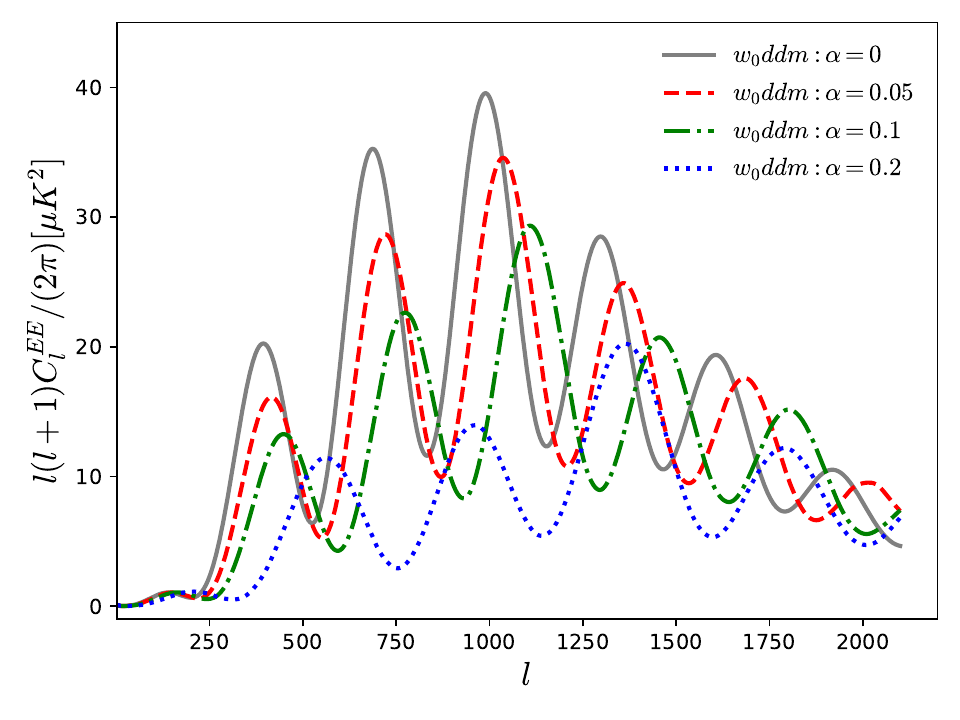}
	\caption{\textit{CMB EE power spectrum obtained by varying $\alpha$  for the $w_0$ddm scenario. } }
	\label{fig:EE-w0ddm}
\end{figure}

In Figs.~\ref{fig:TT-w0ddm}, \ref{fig:TE-w0ddm} and \ref{fig:EE-w0ddm}, we show the effects of the $\alpha$ and $w_0$ parameters in modifying the CMB power spectrum.  We can see that while $w_0$ mainly changes the amplitude of the low-$\ell$ multipoles, $\alpha$ completely modifies the CMB peak structure, changing the peak amplitudes and positions.



\section{Observational data and statistical methodology}
\label{sec-data}
In this section we provide a description of the observational datasets and the statistical methodology. 
To constrain the free parameters of the ddm models we make use of a series of cosmological probes:
\begin{itemize}[noitemsep,topsep=0pt]
\item {\bf CMB}: the Planck 2018 legacy release temperature and polarization CMB measurements~\cite{Aghanim:2018eyx,Aghanim:2019ame}.
\item {\bf lensing}: the Planck 2018 CMB lensing reconstruction likelihood~\cite{Aghanim:2018oex}.
    \item {\bf BAO}: the BAO measurements from 6dFGS~\cite{Beutler:2011hx}, SDSS-MGS~\cite{Ross:2014qpa}, and BOSS DR12~\cite{Alam:2016hwk}. 
    \item {\bf Pantheon}: the Pantheon sample of 1048 Supernovae Type Ia distributed in the redshift interval $z \in [0.01, 2.3]$~\cite{Scolnic:2017caz}.
\item {\bf  R20}: a gaussian prior on the Hubble constant in agreement with the SH0ES collaboration measurement in~\cite{Riess:2020fzl}. 
\end{itemize}

To study the impact of the ddm models on the CMB and LSS observables, we modify the Boltzmann code \texttt{CAMB}~\cite{Lewis:1999bs}. More concretely, we implement the non-standard time evolution of ddm and dr energy densities according to Eqs.~(\ref{eqnar1}) and (\ref{eqnar3}). To investigate and constrain the imprints of ddm models on the CMB and LSS 
we adopt Markov Chain Monte Carlo (MCMC) methods to sample the posterior distribution of the cosmological parameters. We used a modified version of the  MCMC cosmological package  \texttt{CosmoMC}~\cite{Lewis:2002ah,Lewis:2013hha}, publicly available at \url{http://cosmologist.info/cosmomc/}, that is equipped with a convergence diagnostic based on the Gelman-Rubin criterion~\cite{Gelman:1992zz} and supports the Planck 2018 likelihood~\cite{Aghanim:2019ame}. We monitor the convergence of the generated MCMC chains using the standard $R$ parameter, requiring $R - 1 < 0.02$ for the MCMC chains to be considered as converged.  We consider the models outlined in Sec.~\ref{sec-2}, which span three different parameter spaces; namely, a 7-dimensional space for $\Lambda$ddm, 8-dimensional for $w_0$ddm, and 9-dimensional for $w_0w_a$ddm. The free parameters in these models are 
\begin{displaymath}
\mathcal{P}_1 = \{\Omega_{b} h^2, \Omega_{\rm ddm} h^2, 100\theta_{MC}, \tau, n_{s}, {\rm{ln}}(10^{10} A_s), \alpha\},
\end{displaymath}
\begin{displaymath}
\mathcal{P}_2 = \{\Omega_{b} h^2, \Omega_{\rm ddm} h^2, 100\theta_{MC}, \tau, n_{s}, {\rm{ln}}(10^{10} A_s), \alpha, w_0\},
\end{displaymath}
\begin{displaymath}
\mathcal{P}_3 = \{\Omega_{b} h^2, \Omega_{\rm ddm} h^2, 100\theta_{MC}, \tau, n_{s}, {\rm{ln}}(10^{10} A_s), \alpha, w_0, w_a\},
\end{displaymath}
where $\Omega_{b}$ and $\Omega_{\rm ddm}$ are the baryons and ddm densities normalized to the critical density, $\theta_{MC}$  is an approximation of the ratio of sound horizon to the angular diameter distance (which is adopted in \texttt{CosmoMC}~\cite{Lewis:2002ah,Lewis:2013hha} and is based on
fitting formulae given in~\cite{Hu:1995en}), $\tau$ is the reionization optical depth, $n_{s}$ is the scalar spectral index, $A_s$ is the amplitude of the primordial scalar power spectrum, $\alpha$ gauges the strength of the ddm mechanism, and $w_0$ and $w_a$ are the free parameters describing the dark energy equation of state. We impose flat priors on the free parameters as specified in Table~\ref{priors}. 

\begin{table}
\begin{center}
\begin{tabular}{c|c}
\hline
\hline
~~~~~~~~~~Parameter~~~~~~~~~~                    & prior\\
\hline
$\Omega_{b} h^2$         & $[0.005,0.1]$  \\
$\Omega_{\rm ddm} h^2$         & ~~~~~~~~$[0.001,0.99]$~~~~~~~~ \\
$100\theta_{MC}$             & $[0.5,10]$ \\
$\tau$                       & $[0.01,0.8]$ \\
$n_\mathrm{s}$               & $[0.7,1.3]$ \\
${\rm{ln}}(10^{10} A_s)$         & $[1.7, 5.0]$\\
$\alpha$                     & $[0,1]$  \\
$w_0$                        & $[-2, 0.5]$\\
$w_a$                        & $[-3, 3]$\\
\hline 
\hline
\end{tabular}
\end{center}
\caption{Flat priors on various cosmological parameters of the ddm scenarios.}
\label{priors}
\end{table}

\section{Observational Results}
\label{sec-results}

In this section we summarize the observational constraints extracted out of the three distinct ddm scenarios distinguished by the dark energy dependence; namely, {\it (i)}~$\Lambda$ddm, {\it (ii)}~$w_0$ddm and {\it~(iii)}~$w_0w_a$ddm.  

\subsection{ddm and non-dynamical dark energy: The $\bm{\Lambda}$ddm model}

\begin{figure*}
	\centering
	\includegraphics[width=0.72\textwidth]{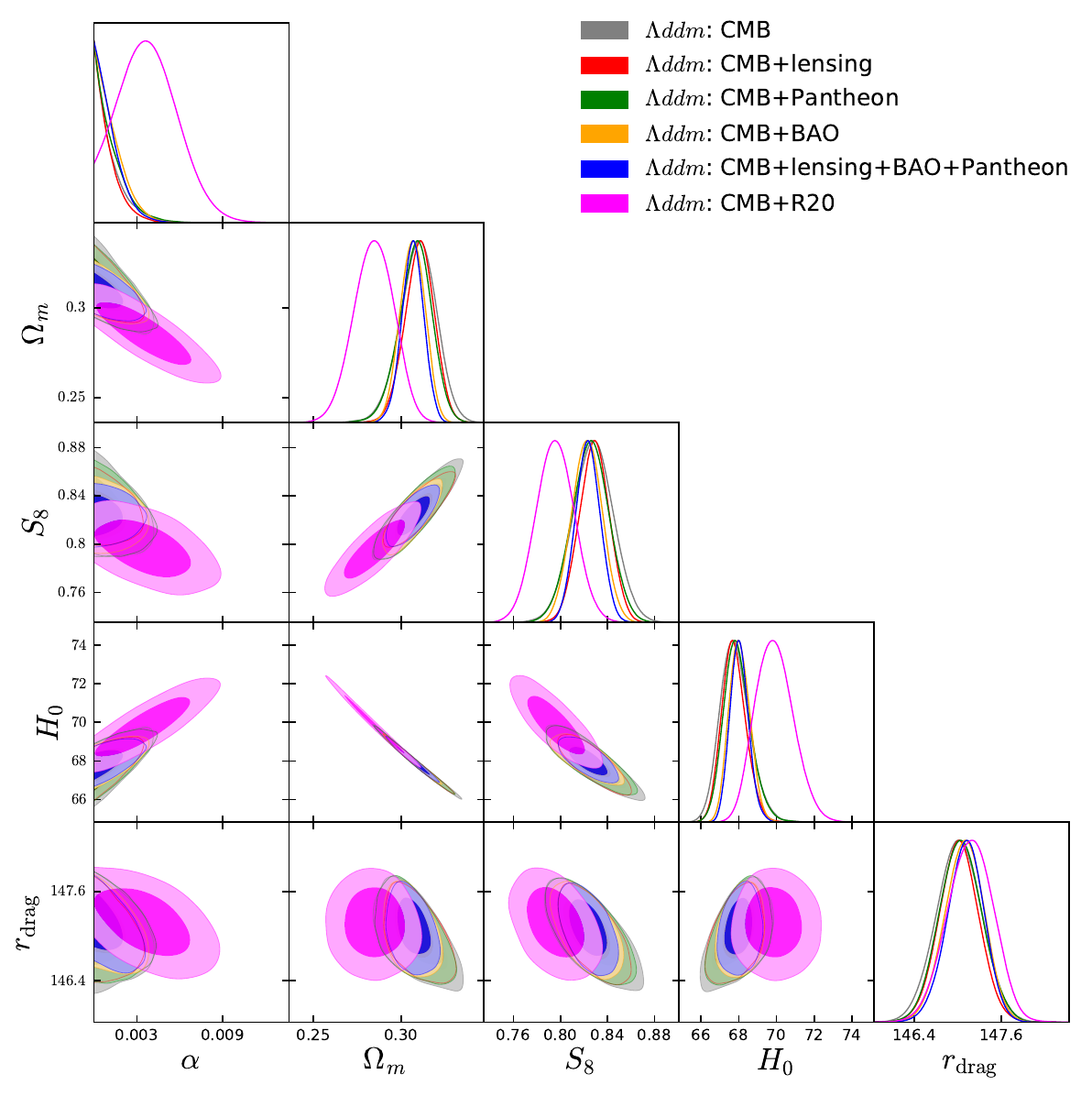}
	\caption{\textit{We show the 1-dimensional posterior distributions of some important parameters and the 2-dimensional joint contours at 68\% and 95\% CL between some of the model parameters  for the $\Lambda$ddm scenario using various cosmological probes and their combinations. $H_0$ and $r_{\rm drag}$ are given in [km/s/Mpc] and [Mpc], respectively.  }}
	\label{fig:LDDM}
\end{figure*}

In Table~\ref{tab:lddm} we display the constraints at 68\% CL and at 95\% CL on the free cosmological parameters of the $\Lambda$ddm model (above the horizontal line) and on the derived parameters (below the horizontal line). In addition, we display the corresponding 2D contour plots in Fig.~\ref{fig:LDDM}. Remarkably, independently of the dataset combinations considered, we obtain an upper limit for $\alpha$  that is consistent with zero, recovering the constraints on the cosmological parameters that we have for a standard $\Lambda$CDM model,  and therefore the same level of tension with $H_0$ and $S_8$. Moreover, the upper bound on $\alpha$ is robust and does not change among the various dataset combinations, with the only exception being the CMB+lensing dataset, for which the bound becomes more restrictive by a factor of $0.8$,  and when the R20 prior is included, as discussed in the Appendix. In particular, $\alpha<0.0033$ at 95\% CL for CMB alone, whereas $\alpha<0.0027$ at 95\% CL for CMB+lensing, while it remains unaltered when BAO or Pantheon are included in the analysis.

We can conclude that while the analysis of Planck 2015 (high-$\ell$ TT + low-$\ell$
TEB)+R18+JLA+BAO data suggests that the $\Lambda$ddm model could help to reduce (though not fully eliminate) the $H_0$ and $S_8$
 tensions (specifically, a discrepancy persists at $2.5\sigma$ and $1.5\sigma$, respectively)~\cite{Pandey:2019plg}, the addition of the
high-$\ell$ polarization data of Planck 2018 into the analysis changes the overall picture: as  we have shown in Table~\ref{tab:lddm} {\it the $\Lambda$ddm model cannot solve the $H_0$ tension below  $3\sigma$. }

From Fig.~\ref{fig:LDDM} it is possible to understand the effect of the $\Lambda$ddm model on the $H_0$ and $S_8$ tensions analyzing the correlation between $\alpha$ and the relevant cosmological parameters. The correlation works in the right direction to potentially alleviate the current cosmological tensions, being positively correlated with $H_0$ (the Hubble value shifts about $1\sigma$ towards a higher value) and at the same time inversely correlated with the sound horizon $r_{\rm drag}$ (although mildly in this case), and anticorrelated with both $S_8$ and $\Omega_m$.  Therefore, we can speculate that even if the observational datasets do not favor  
the $\Lambda$ddm scenario for the present interaction rate, however, the picture may change for a different interaction rate. As mentioned in section \ref{sec-2}, since the choice of the interaction rate is not unique here and one may consider a very general interaction rate of the form  $Q \equiv \mathcal{H} Q (\rho_{\rm ddm}, \rho_{\rm dr})$ or $Q  \equiv Q (\rho_{\rm ddm}, \rho_{\rm dr})$, therefore, an interaction rate other than $Q \propto \mathcal{H} \rho_{\rm ddm}$ could offer different constraints on the free and derived parameters.
\begingroup                                                                                                                     
\squeezetable                                                                                                                   
\begin{center}                                                                                                                  
\begin{table*}                                                                                                                    \resizebox{\textwidth}{!}{                        
\begin{tabular}{cccccccc}                                                                                                            
\hline\hline                                                                                                                    
Parameters & CMB & CMB+lensing & CMB+Pantheon& CMB+BAO & CMB+lensing+Pantheon+BAO & CMB+R20 \\ \hline
$\Omega_{\rm ddm} h^2$ & $    0.1194_{-    0.0015-    0.0032}^{+    0.0016+    0.0030}$ & $    0.1195_{-    0.0013-    0.0025}^{+    0.0013+    0.0025}$& $    0.1191_{-    0.0014-    0.0029}^{+    0.0015+    0.0027}$  & $    0.1189_{-    0.0011-    0.0021}^{+    0.0011+    0.0021}$ & $    0.11889_{-    0.00094-    0.0019}^{+    0.00094+    0.0019}$ & $    0.1159_{-    0.0016-    0.0033}^{+    0.0016+    0.0031}$\\

$\Omega_b h^2$ & $    0.02231_{-    0.00015-    0.00031}^{+    0.00015+    0.00030}$ & $    0.02233_{-    0.00014-    0.00028}^{+    0.00014+    0.00028}$ & $    0.02232_{-    0.00015-    0.00030}^{+    0.00015+    0.00030}$ & $    0.02234_{-    0.00015-    0.00030}^{+    0.00015+    0.00030}$ & $    0.02235_{-    0.00014-    0.00029}^{+    0.00015+    0.00029}$  & $    0.02235_{-    0.00017-    0.00036}^{+    0.00017+    0.00035}$ \\

$100\theta_{MC}$ & $    1.04092_{-    0.00031-    0.00061}^{+    0.00031+    0.00061}$ & $    1.04091_{-    0.00030-    0.00060}^{+    0.00030+    0.00060}$ & $    1.04093_{-    0.00030-    0.00060}^{+    0.00031+    0.00061}$ & $    1.04097_{-    0.00029-    0.00057}^{+    0.00029+    0.00058}$  & $    1.04097_{-    0.00029-    0.00058}^{+    0.00029+    0.00058}$ & $    1.04118_{-    0.00030-    0.00059}^{+    0.00030+    0.00060}$\\

$\tau$ & $    0.0546_{-    0.0081-    0.015}^{+    0.0074+    0.016}$ & $    0.0550_{-    0.0076-    0.015}^{+    0.0072+    0.015}$ & $    0.0550_{-    0.0076-    0.015}^{+    0.0077+    0.016}$ & $    0.0552_{-    0.0080-    0.015}^{+    0.0073+    0.016}$ & $    0.0565_{-    0.0070-    0.014}^{+    0.0070+    0.015}$  & $    0.0578_{-    0.0083-    0.015}^{+    0.0076+    0.017}$ \\

$n_s$ & $    0.9653_{-    0.0045-    0.0087}^{+    0.0045+    0.0089}$ & $    0.9652_{-    0.0041-    0.0079}^{+    0.0041+    0.0080}$ & $    0.9658_{-    0.0041-    0.0082}^{+    0.0041+    0.0082}$ & $    0.9664_{-    0.0038-    0.0077}^{+    0.0040+    0.0076}$ & $    0.9665_{-    0.0037-    0.0073}^{+    0.0038+    0.0073}$  & $    0.9708_{-    0.0041-    0.0081}^{+    0.0042+    0.0080}$ \\

${\rm{ln}}(10^{10} A_s)$ & $    3.046_{-    0.015-    0.031}^{+    0.016+    0.032}$ & $    3.047_{-    0.015-    0.029}^{+    0.014+    0.029}$ & $    3.047_{-    0.016-    0.032}^{+    0.016+    0.033}$ & $    3.046_{-    0.016-    0.030}^{+    0.016+    0.032}$ & $    3.049_{-    0.014-    0.028}^{+    0.014+    0.029}$  & $    3.051_{-    0.016-    0.031}^{+    0.016+    0.033}$ \\

$\alpha$ & $ <0.0013\,<0.0033 $ & $  <0.0011\,<0.0027$  & $  <0.0015\,<0.0033 $ & $   <0.0015\,<0.0031 $  & $ <0.0013\,<0.0029 $  & $ 0.0038^{+0.0019}_{-0.0023}\, <  0.0073 $ \\

\hline
$\Omega_m$ & $    0.310_{-    0.010-    0.021}^{+    0.011+    0.020}$ & $    0.3105_{-    0.0081-    0.017}^{+    0.0089+    0.016}$ & $    0.308_{-    0.009-    0.020}^{+    0.010+    0.017}$ & $    0.3066_{-    0.0071-    0.014}^{+    0.0071+    0.013}$ & $    0.3066_{-    0.0061-    0.013}^{+    0.0062+    0.012}$  & $    0.284_{-    0.011-    0.022}^{+    0.011+    0.021}$ \\

$\sigma_8$ & $    0.8149_{-    0.0077-    0.015}^{+    0.0077+    0.016}$ & $    0.8147_{-    0.0071-    0.013}^{+    0.0063+    0.014}$ & $    0.8146_{-    0.0085-    0.016}^{+    0.0078+    0.017}$ & $    0.8138_{-    0.0084-    0.015}^{+    0.0077+    0.016}$ & $    0.8145_{-    0.0074-    0.013}^{+    0.0065+    0.014}$ & $    0.817_{-    0.010-    0.018}^{+    0.009+    0.019}$ \\

$H_0\,[{\rm km/s/Mpc}]$ & $   67.81_{-    0.89-    1.5}^{+    0.68+    1.7}$ & $   67.75_{-    0.70-    1.2}^{+    0.59+    1.3}$ & $   67.93_{-    0.81-    1.4}^{+    0.62+    1.5}$ & $   68.04_{-    0.60-    1.1}^{+    0.53+    1.1}$ & $   68.03_{-    0.52-    0.99}^{+    0.46+    0.98}$  & $   69.9_{-    1.1-    1.9}^{+    0.9+    2.0}$ \\

$S_8$ & $    0.828_{-    0.017-    0.034}^{+    0.017+    0.033}$  & $    0.829_{-    0.013-    0.025}^{+    0.013+    0.025}$ & $    0.825_{-    0.016-    0.031}^{+    0.016+    0.030}$ & $    0.823_{-    0.013-    0.025}^{+    0.013+    0.025}$ & $    0.823_{-    0.010-    0.020}^{+    0.010+    0.020}$  & $    0.796_{-    0.016-    0.031}^{+    0.016+    0.032}$ \\

$r_{\rm{drag}}\,[{\rm Mpc}]$ & $  147.02_{-    0.31-    0.60}^{+    0.30+    0.59}$ & $  147.01_{-    0.27-    0.54}^{+    0.27+    0.53}$ & $  147.04_{-    0.29-    0.56}^{+    0.29+    0.58}$ & $  147.09_{-    0.27-    0.54}^{+    0.27+    0.51}$ & $  147.10_{-    0.24-    0.52}^{+    0.27+    0.48}$  & $  147.17_{-    0.31-    0.62}^{+    0.32+    0.59}$ \\

\hline\hline                                                                                                                    
\end{tabular}}                                           
\caption{68\% and 95\% CL on the free cosmological parameters of the $\Lambda$ddm scenario (above the line) and the derived ones (below the line). }\label{tab:lddm} 

\end{table*}                                                                                                                     
\end{center}                                                                                                                    
\endgroup 


\begingroup                                                                                                                     
\squeezetable                                                                                                                   
\begin{center}                                                                                                                  
\begin{table*}                                                                                                                    \resizebox{\textwidth}{!}{                        
\begin{tabular}{ccccccccccc}                                                                                                            
\hline\hline                                                                                                                    
Parameters & CMB & CMB+lensing & CMB+Pantheon& CMB+BAO & CMB+lensing+Pantheon+BAO & CMB+R20 \\ \hline
$\Omega_{\rm ddm} h^2$ & $    0.1191_{-    0.0015-    0.0032}^{+    0.0016+    0.0029}$ & $    0.1186_{-    0.0013-    0.0028}^{+    0.0015+    0.0026}$& $    0.1195_{-    0.0014-    0.0032}^{+    0.0017+    0.0029}$  & $    0.1191_{-    0.0014-    0.0029}^{+    0.0014+    0.0028}$ & $    0.1192_{-    0.0011-    0.0022}^{+    0.0011+    0.0022}$ & $    0.1194_{-    0.0015-    0.0030}^{+    0.0015+    0.0029}$  \\

$\Omega_b h^2$ & $    0.02234_{-    0.00016-    0.00030}^{+    0.00016+    0.00031}$ & $    0.02236_{-    0.00015-    0.00031}^{+    0.00016+    0.00030}$ & $    0.02231_{-    0.00015-    0.00030}^{+    0.00015+    0.00030}$ & $    0.02233_{-    0.00015-    0.00030}^{+    0.00015+    0.00029}$ & $    0.02235_{-    0.00015-    0.00029}^{+    0.00015+    0.00029}$ & $    0.02232_{-    0.00015-    0.00030}^{+    0.00015+    0.00028}$ \\

$100\theta_{MC}$ & $    1.04096_{-    0.00031-    0.00062}^{+    0.00030+    0.00061}$ & $    1.04098_{-    0.00031-    0.00062}^{+    0.00031+    0.00060}$ & $    1.04090_{-    0.00031-    0.00060}^{+    0.00031+    0.00062}$ & $    1.04093_{-    0.00031-    0.00059}^{+    0.00031+    0.00058}$  & $    1.04094_{-    0.00030-    0.00057}^{+    0.00030+    0.00058}$ & $    1.04091_{-    0.00030-    0.00061}^{+    0.00031+    0.00060}$\\

$\tau$ & $    0.0542_{-    0.0073-    0.015}^{+    0.0074+    0.015}$ & $    0.0533_{-    0.0074-    0.015}^{+    0.0074+    0.015}$ & $    0.0546_{-    0.0081-    0.015}^{+    0.0073+    0.016}$ & $    0.0547_{-    0.0082-    0.015}^{+    0.0073+    0.016}$ & $    0.0554_{-    0.0079-    0.014}^{+    0.0068+    0.015}$ & $    0.0546_{-    0.0080-    0.015}^{+    0.0075+    0.016}$\\

$n_s$ & $    0.9659_{-    0.0043-    0.0086}^{+    0.0043+    0.0087}$ & $    0.9667_{-    0.0042-    0.0083}^{+    0.0042+    0.0085}$ & $    0.9651_{-    0.0044-    0.0084}^{+    0.0043+    0.0084}$ & $    0.9659_{-    0.0041-    0.0082}^{+    0.0041+    0.0085}$ & $    0.9657_{-    0.0039-    0.0075}^{+    0.0038+    0.0075}$ & $    0.9652_{-    0.0043-    0.0085}^{+    0.0042+    0.0086}$\\

${\rm{ln}}(10^{10} A_s)$ & $    3.045_{-    0.015-    0.029}^{+    0.015+    0.031}$ & $    3.042_{-    0.015-    0.029}^{+    0.014+    0.030}$ & $    3.046_{-    0.015-    0.031}^{+    0.016+    0.032}$ & $    3.046_{-    0.017-    0.031}^{+    0.015+    0.033}$ & $    3.047_{-    0.016-    0.027}^{+    0.014+    0.030}$  & $    3.046_{-    0.016-    0.032}^{+    0.016+    0.033}$\\

$\alpha$ & $ <0.0014\,<0.0035 $ & $  <0.0015\,<0.0035$  & $  <0.0013\,<0.0032 $ & $   <0.0014\,<0.0031 $  & $ <0.0012\,<0.0027 $  & $  < 0.0012 < 0.0030 $  \\

$w_0$ & $ -1.58_{-0.34-0.41}^{+0.15+0.51} $ & $  -1.57_{-0.32-0.39}^{+0.15+0.49}$  & $  -1.026_{-0.037-0.073}^{+0.038+0.075} $ & $   -1.020_{-0.056-0.12}^{+0.065+0.011} $  & $ -1.019_{-0.031-0.063}^{+0.032+0.062} $  & $   -1.182_{-    0.049-    0.096}^{+    0.050+    0.096}$ \\
\hline

$\Omega_m$ & $    0.193_{-    0.055-    0.067}^{+    0.018+    0.097}$ & $    0.191_{-    0.053-    0.065}^{+    0.017+    0.095}$ & $    0.304_{-    0.011-    0.022}^{+    0.011+    0.023}$ & $    0.303_{-    0.012-    0.024}^{+    0.012+    0.023}$ & $    0.3036_{-    0.0075-    0.015}^{+    0.0076+    0.015}$  & $    0.264_{-    0.011-    0.019}^{+    0.009+    0.020}$\\

$\sigma_8$ & $    0.977_{-    0.044-    0.15}^{+    0.094+    0.12}$ & $    0.973_{-    0.041-    0.14}^{+    0.089+    0.011}$ & $    0.823_{-    0.014-    0.027}^{+    0.014+    0.027}$ & $    0.820_{-    0.021-    0.037}^{+    0.019+    0.039}$ & $    0.820_{-    0.011-    0.021}^{+    0.011+    0.022}$  & $    0.866_{-    0.016-    0.031}^{+    0.016+    0.032}$\\

$H_0\,[{\rm km/s/Mpc}]$ & $   87_{-    5-    18}^{+    12+    15}$ & $   88_{-    5-    18}^{+    12+    14}$ & $   68.5_{-    1.1-    2.2}^{+    1.1+    2.2}$ & $   68.5_{-    1.6-    2.6}^{+    1.3+    3.0}$ & $   68.45_{-    0.81-    1.6}^{+    0.81+    1.6}$ & $   73.4_{-    1.3-    2.6}^{+    1.3+    2.5}$\\

$S_8$ & $    0.772_{-    0.036-    0.053}^{+    0.023+    0.060}$  & $    0.766_{-    0.036-    0.050}^{+    0.020+    0.060}$ & $    0.827_{-    0.016-    0.032}^{+    0.016+    0.032}$ & $    0.824_{-    0.013-    0.027}^{+    0.013+    0.026}$ & $    0.824_{-    0.010-    0.020}^{+    0.010+    0.021}$ & $    0.813_{-    0.015-    0.030}^{+    0.015+    0.030}$ \\

$r_{\rm{drag}}\,[{\rm Mpc}]$ & $  147.02_{-    0.30-    0.58}^{+    0.30+    0.59}$ & $  147.12_{-    0.28-    0.54}^{+    0.28+    0.55}$ & $  146.98_{-    0.30-    0.58}^{+    0.30+    0.60}$ & $  147.05_{-    0.29-    0.56}^{+    0.28+    0.57}$ & $  147.06_{-    0.25-    0.51}^{+    0.26+    0.49}$  & $  147.00_{-    0.29-    0.57}^{+    0.29+    0.57}$  \\
\hline\hline                                                                                                                    
\end{tabular}          }                                    
\caption{68\% and 95\% CL on the free cosmological parameters of the $w_0$ddm model (above the horizontal line) and the derived ones (below the horizontal line).}\label{tab:wddm}                              
\end{table*}                              
\end{center}                                              
\endgroup 
\begin{figure*}
	\centering
	\includegraphics[width=0.7\textwidth]{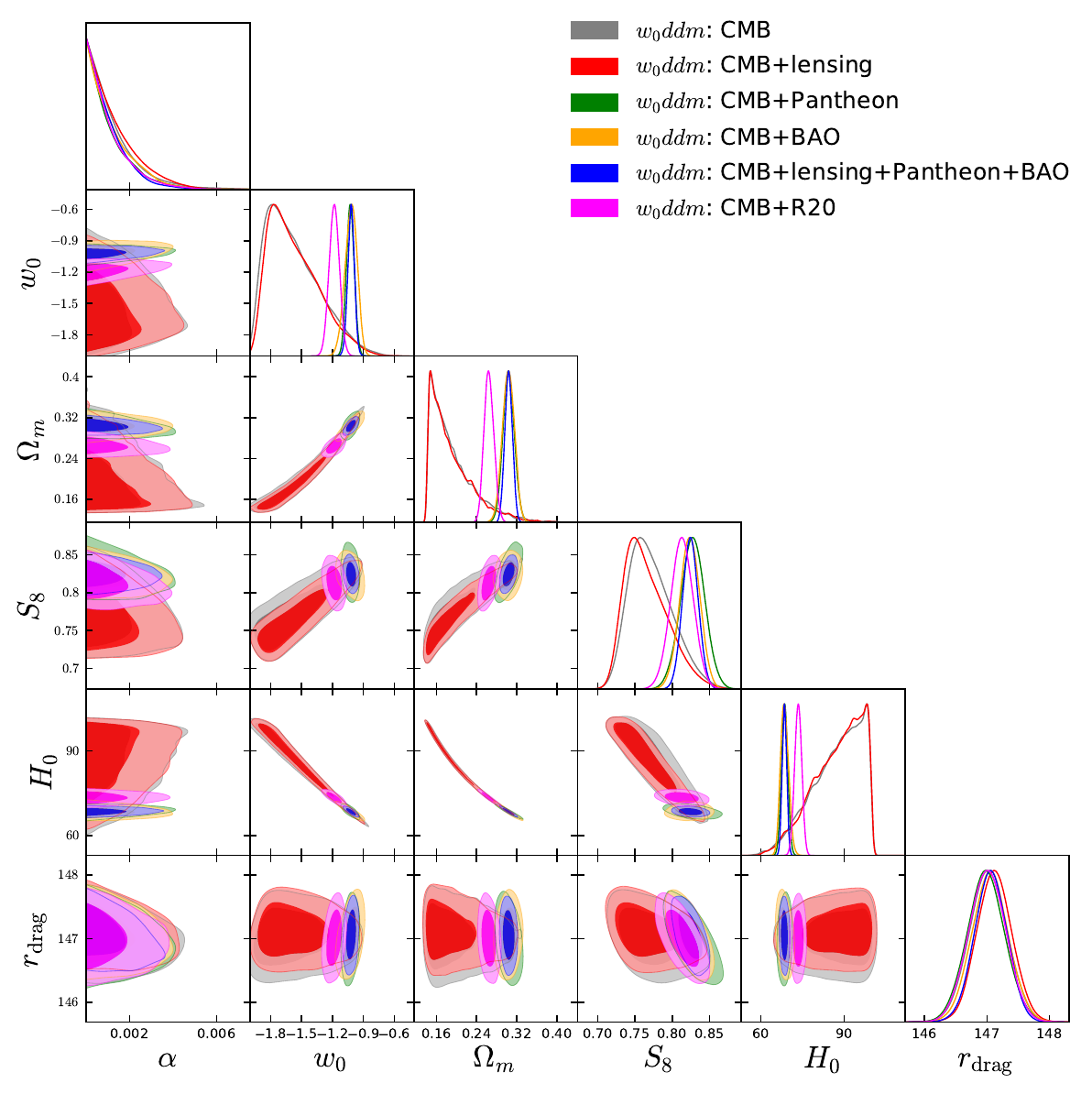}
	\caption{\textit{We show the 1-dimensional posterior distributions of some important parameters and the 2-dimensional joint contours at 68\% and 95\% CL between some of the model parameters  for the $w_0$ddm scenario using various cosmological probes and their combinations. $H_0$ and $r_{\rm drag}$ are given in [km/s/Mpc] and [Mpc], respectively.  }}
	\label{fig:wDDM}
\end{figure*}
\begingroup                                                                                                                     
\squeezetable                                                                                                                   
\begin{center}                                                                                                                  
\begin{table*}                                                                                                                    \resizebox{\textwidth}{!}{                        
\begin{tabular}{ccccccccccc}                                                                                                            
\hline\hline                                                                                                                    
Parameters & CMB & CMB+lensing & CMB+Pantheon& CMB+BAO & CMB+lensing+Pantheon+BAO & CMB+R20 \\ \hline
$\Omega_{\rm ddm} h^2$ & $    0.1189_{-    0.0015-    0.0032}^{+    0.0016+    0.0030}$ & $    0.1185_{-    0.0013-    0.0028}^{+    0.0015+    0.0027}$& $    0.1195_{-    0.0014-    0.0032}^{+    0.0017+    0.0030}$  & $    0.1196_{-    0.0014-    0.0027}^{+    0.0014+    0.0027}$ & $    0.1195_{-    0.0011-    0.0023}^{+    0.0012+    0.0022}$ & $    0.1191_{-    0.0015-    0.0032}^{+    0.0016+    0.0029}$\\

$\Omega_b h^2$ & $    0.02235_{-    0.00016-    0.00031}^{+    0.00016+    0.00031}$ & $    0.02238_{-    0.00016-    0.00031}^{+    0.00016+    0.00031}$ & $    0.02232_{-    0.00016-    0.00030}^{+    0.00015+    0.00030}$ & $    0.02232_{-    0.00015-    0.00029}^{+    0.00015+    0.00030}$ & $    0.02232_{-    0.00015-    0.00029}^{+    0.00014+    0.00028}$ & $    0.02234_{-    0.00015-    0.00031}^{+    0.00015+    0.00031}$  \\

$100\theta_{MC}$ & $    1.04096_{-    0.00032-    0.00063}^{+    0.00031+    0.00063}$ & $    1.04100_{-    0.00030-    0.00058}^{+    0.00030+    0.00061}$ & $    1.04089_{-    0.00032-    0.00065}^{+    0.00033+    0.00064}$ & $    1.04089_{-    0.00030-    0.00063}^{+    0.00033+    0.00060}$  & $    1.04091_{-    0.00030-    0.00059}^{+    0.00030+    0.00061}$  & $    1.04093_{-    0.00031-    0.00059}^{+    0.00031+    0.00061}$ \\

$\tau$ & $    0.0544_{-    0.0078-    0.016}^{+    0.0079+    0.016}$ & $    0.0530_{-    0.0074-    0.015}^{+    0.0073+    0.016}$ & $    0.0542_{-    0.0074-    0.015}^{+    0.0075+    0.016}$ & $    0.0539_{-    0.0078-    0.015}^{+    0.0076+    0.016}$ & $    0.0544_{-    0.0080-    0.015}^{+    0.0074+    0.016}$ & $    0.0545_{-    0.0082-    0.015}^{+    0.0074+    0.016}$  \\

$n_s$ & $    0.9662_{-    0.0044-    0.0088}^{+    0.0043+    0.0084}$ & $    0.9672_{-    0.0040-    0.0081}^{+    0.0040+    0.0081}$ & $    0.9654_{-    0.0047-    0.0085}^{+    0.0044+    0.0088}$ & $    0.9649_{-    0.0041-    0.0084}^{+    0.0045+    0.0081}$ & $    0.9652_{-    0.0040-    0.0079}^{+    0.0040+    0.0077}$ & $    0.9658_{-    0.0043-    0.0084}^{+    0.0043+    0.0085}$ \\

${\rm{ln}}(10^{10} A_s)$ & $    3.045_{-    0.016-    0.033}^{+    0.016+    0.033}$ & $    3.040_{-    0.015-    0.029}^{+    0.014+    0.031}$ & $    3.045_{-    0.015-    0.032}^{+    0.015+    0.032}$ & $    3.045_{-    0.016-    0.032}^{+    0.016+    0.032}$ & $    3.045_{-    0.015-    0.030}^{+    0.015+    0.031}$ & $    3.045_{-    0.016-    0.031}^{+    0.016+    0.031}$  \\

$\alpha$ & $ <0.0015\,<0.0035 $ & $  <0.0015\,<0.0035$  & $  <0.0012\,<0.0031 $ & $   <0.0012\,<0.0028 $  & $ <0.0011\,<0.0026 $ & $  < 0.0014 < 0.0033 $  \\

$w_0$ & $ -1.17_{-0.58}^{+0.33} <-0.28 $ & $  -1.18_{-0.56}^{+0.37}<-0.34$  & $  -0.88_{-0.14-0.25}^{+0.13+0.26} $ & $   -0.59_{-0.27-0.53}^{+0.27+0.51} $  & $ -0.962_{-0.078-0.15}^{+0.078+0.16} $  & $   -0.81_{-    0.16-    0.51}^{+    0.32+    0.41}$  \\

$w_a$ & $ <-0.89\,<0.64 $ & $  <-0.81<0.73$  & $  -0.71_{-0.58-1.3}^{+0.73+1.2} $ & $   -1.19_{-0.75-1.5}^{+0.80+1.4} $  & $ -0.24_{-0.27-0.60}^{+0.32+0.58} $  & $   <-1.15\,<0.30$  \\
\hline
$\Omega_m$ & $    0.214_{-    0.078-    0.10}^{+    0.022+    0.15}$ & $    0.211_{-    0.073-    0.09}^{+    0.023+    0.13}$ & $    0.295_{-    0.014-    0.025}^{+    0.013+    0.028}$ & $    0.339_{-    0.026-    0.049}^{+    0.026+    0.051}$ & $    0.3044_{-    0.0076-    0.015}^{+    0.0077+    0.015}$ & $    0.265_{-    0.010-    0.018}^{+    0.009+    0.020}$  \\

$\sigma_8$ & $    0.949_{-    0.05-    0.18}^{+    0.12+    0.15}$ & $    0.946_{-    0.51-    0.17}^{+    0.11+    0.13}$ & $    0.834_{-    0.016-    0.034}^{+    0.018+    0.031}$ & $    0.794_{-    0.026-    0.045}^{+    0.023+    0.049}$ & $    0.822_{-    0.011-    0.022}^{+    0.011+    0.022}$  & $    0.868_{-    0.016-    0.032}^{+    0.017+    0.033}$  \\

$H_0\,[{\rm km/s/Mpc}]$ & $   84_{-    7-    22}^{+    15+    18}$ & $   84_{-    7-    21}^{+    14+    17}$ & $   69.5_{-    1.4-    2.8}^{+    1.4+    2.8}$ & $   65.0_{-    2.8-    4.8}^{+    2.2+    5.0}$ & $   68.43_{-    0.81-    1.6}^{+    0.81+    1.6}$ & $   73.3_{-    1.3-    2.5}^{+    1.3+    2.6}$ \\

$S_8$ & $    0.783_{-    0.046-    0.063}^{+    0.025+    0.077}$  & $    0.776_{-    0.045-    0.059}^{+    0.024+    0.071}$ & $    0.827_{-    0.016-    0.033}^{+    0.017+    0.032}$ & $    0.843_{-    0.018-    0.036}^{+    0.018+    0.034}$ & $    0.828_{-    0.011-    0.022}^{+    0.011+    0.022}$ & $    0.815_{-    0.015-    0.029}^{+    0.015+    0.030}$ \\

$r_{\rm{drag}}\,[{\rm Mpc}]$ & $  147.05_{-    0.30-    0.61}^{+    0.30+    0.60}$ & $  147.15_{-    0.27-    0.56}^{+    0.27+    0.54}$ & $  146.99_{-    0.29-    0.58}^{+    0.29+    0.62}$ & $  146.98_{-    0.29-    0.57}^{+    0.29+    0.59}$ & $  147.03_{-    0.27-    0.54}^{+    0.27+    0.53}$ & $  147.03_{-    0.29-    0.57}^{+    0.30+    0.59}$   \\
\hline\hline                                                                                                                    
\end{tabular}}                                            
\caption{68\% and 95\% CL on the free cosmological parameters of the $w_0w_a$ddm model (above the horizontal line) and the derived ones (below the horizontal line). }\label{tab:w0waddm}
\end{table*}                                                 
\end{center}                                            
\endgroup 
\begin{figure*}
	\centering
	\includegraphics[width=0.7\textwidth]{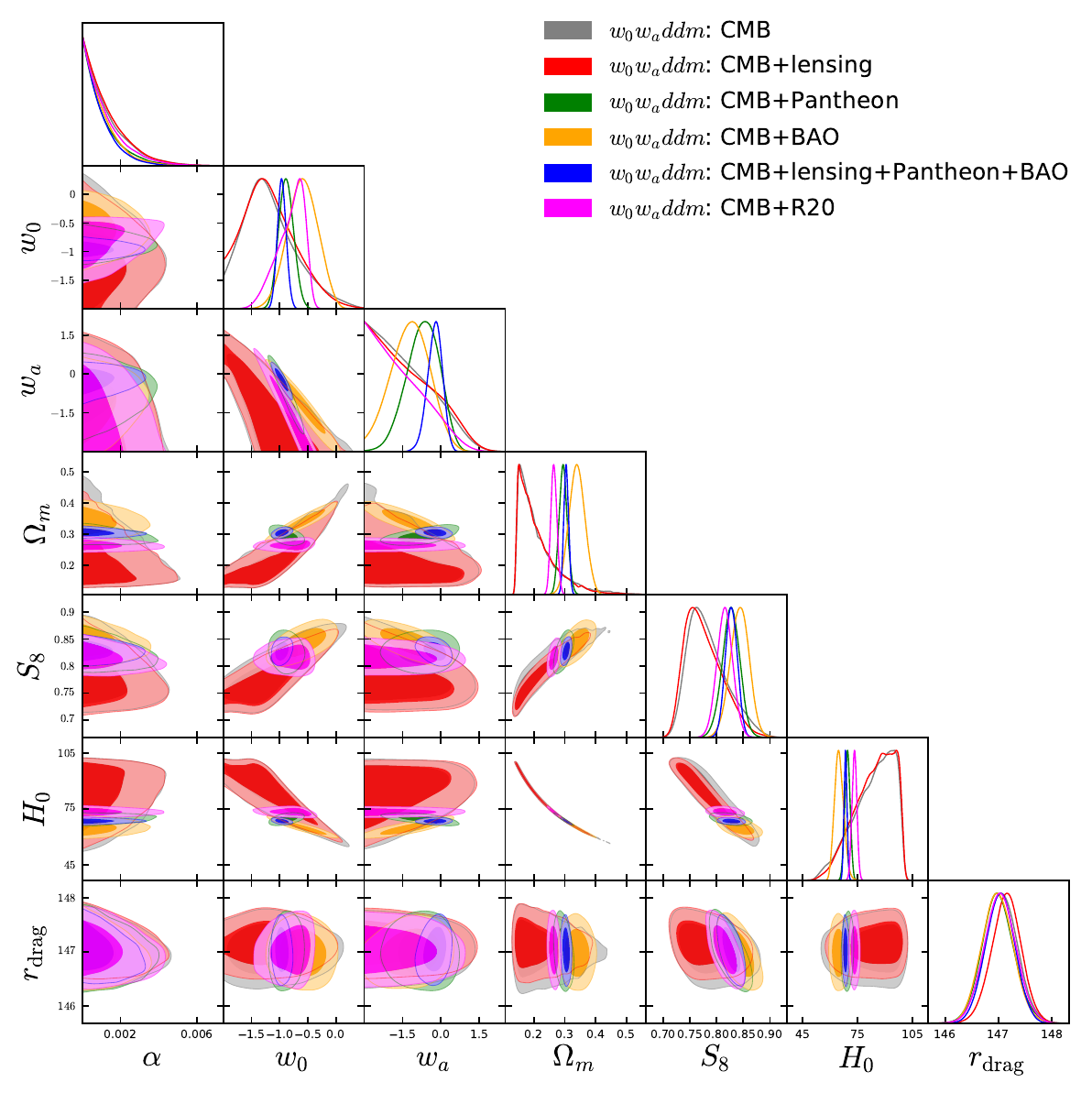}
	\caption{\textit{We show the 1-dimensional posterior distributions of some important parameters and the 2-dimensional joint contours at 68\% and 95\% CL between some of the model parameters  for the $w_0w_a$ddm scenario using various cosmological probes and their combinations. $H_0$ and $r_{\rm drag}$ are given in [km/s/Mpc] and [Mpc], respectively.}}
	\label{fig:w0waDDM}
\end{figure*}

\subsection{ddm and dynamical dark energy:  The $\bm{w_0}$ddm and $\bm{w_0w_a}$ddm models} 

It is well-known that both early-time models (departing from $\Lambda$CDM before recombination) and late-time models (departing from $\Lambda$CDM after recombination) cannot alleviate the $H_0$ tension if taken separately. Therefore, in this section we combine the most simple late-time models featuring a dark energy equation of state free to vary with the proposed early-time ddm model and investigate whether a correlation between parameters of this hybrid set up could play a role in the determination of the $\alpha$ parameter.

We begin by considering an extended set up in which the DE equation of state is a constant, $w_0$, and hence it is expected that the effects of the dynamical dark energy can be realized from the behaviour of this ddm scenario.  In Table~\ref{tab:wddm} we present the constraints at 68\% CL and 95\% CL on the cosmological parameters of the $w_0$ddm scenario. The corresponding 1D posterior distributions and 2D contour plots are shown in Fig.~\ref{fig:wDDM}. 

As in our previous analysis a robust upper limit on $\alpha$ is obtained for all the dataset combinations,  also when the R20 prior is included. Actually, the upper limits on $\alpha$ in this scenario are almost similar to those bounding the $\Lambda$ddm model. Moreover, since $\alpha$ is consistent with zero, the best-fit values of the cosmological parameters are similar to those constraining the $w_0$CDM scenario; see e.g., Table~II of Ref.~\cite{Anchordoqui:2021gji}.  

 Since this $w_0$ddm scenario solves both the $H_0$ and $S_8$ tensions, we can safely combine Planck and the R20 prior to understand if the resolution of the Hubble tension emerges at the price of $\alpha\neq0$ or a phantom dark energy $w_0<-1$. We conclude that the solution is due to the latter because $\alpha$ remains consistent with zero. We can see that the combination CMB+R20 fixes $H_0$ in agreement with SH0ES at $1\sigma$ level, but the replacement of $w_0$ by $\Lambda$ exacerbates the $S_8$ tension. Moreover, as we can see in Fig.~\ref{fig:wDDM}, the introduction of $w_0$ breaks down the correlation between $\alpha$ and the parameters of interest, such as $H_0$, $r_{\rm drag}$, $S_8$, and $\Omega_m$.

Next, we consider a model in which the dark energy equation of state has a dynamical nature in terms of the CPL parametrization. In Table~\ref{tab:w0waddm} we show the constraints at 68\% CL and 95\% CL on the cosmological parameters of this scenario. The corresponding 2D contour plots are shown in Fig.~\ref{fig:w0waDDM}. 
This further extension of the dynamical DE sector does not change the conclusions of the previous analyses: $\alpha$ does not correlate with the other parameters of the model, and we have a robust upper limit on $\alpha$ for all the dataset combinations. The constraints on the cosmological parameters are similar to those obtained in the $w_0w_a$CDM model, because $\alpha$ is always consistent with zero.

The absence of correlation between $\alpha$ and the dark energy equation of state can be deduced by comparing the upper and bottom panels of Fig.~\ref{fig:TT-w0ddm}. This is because even if the effects of $\alpha$ and $w_0$ are similar in the low-$\ell$ multipole range, contrarily to $w_0$, $\alpha$ affects strongly the position of the peaks and the smoothing of the damping tail. Since this high multipole region is extremely well constrained by the peak structure of the CMB data, the addition of external datasets or extensions of the minimal $\Lambda$ddm model do not affect the upper bound on $\alpha$.


\section{Conclusions}
\label{sec-concl}

We have investigated a specific cosmological set-up where the CDM sector is disintegrated into dark radiation in the background of a homogeneous and isotropic universe and the remaining matter components do not take part in this dynamics. The dark energy sector has been assumed to be either non-dynamical (the cosmological constant) or dynamical (characterized by some equation of state). 
We considered a typical functional form of the disintegration rate $Q = \alpha {\cal H} \rho_{\rm ddm}$, where $\alpha > 0$  describes the strength of the disintegration~\cite{Bjaelde:2012wi}. The non-dynamical model has been proposed as a way of simultaneously diminishing both the $H_0$ and the $S_8$ tensions, but not to fully resolve them~\cite{Pandey:2019plg}. This proposal is rooted in the analysis of Planck 2015 (high-$\ell$TT+low-$\ell$TEB)+R18+JLA+BAO data. A point worth noting at this juncture is that adopting the $3^{\rm rd}$ criterion of the $H_0$ Olympics  one can indeed venture to argue that the $\Lambda$ddm model could help to reduce (though not fully eliminate) the $H_0$ and $S_8$ tensions.\footnote{The $H_0$ Olympics
  establishes a set of criteria to compare the relative success of proposed models to address the $H_0$ tension~\cite{Schoneberg:2021qvd}.} Specifically, when  $\Lambda$ddm is confronted with the above mentioned data a discrepancy persists at $2.5\sigma$ with the SH0ES $H_0$ measurement and at $1.5\sigma$ with the local determination of the weighted amplitude of matter fluctuations, and where the $\chi^2$ of the
combined fits to Planck2015 (high-$\ell$TT+low-$\ell$TEB)+JLA+BAO+R18 data are 11978.3 and 11977.5 for
 $\Lambda$CDM and $\Lambda$ddm, respectively~\cite{Pandey:2019plg}.  We have shown herein that the addition of the
high-$\ell$ polarization data of Planck 2018 into the analysis acutely
changes the picture as the $\Lambda$ddm model cannot solve the $H_0$
tension below the standard $3\sigma$ exclusion level adopted in the
$H_0$ Olympics; see Table~\ref{tab:lddm} and Fig.~\ref{fig:LDDM}.

In a second phase of the investigation we studied two related models endowed with dynamical dark energy $-$ one in which the dark energy equation of state is constant with redshift, $w_{\rm de} (a) = w_0$, and another where $w_{\rm de} (a)$ assumes the CPL parametrization $w_{\rm de} (a) = w_0 + w_a (1-a)$. We performed Monte Carlo Markov chain analyses for different combinations of data sets and for different parameter sets. The results of the second-phase data-analysis are encapsulated in Tables~\ref{tab:wddm} (for $w_0$ddm) and  \ref{tab:w0waddm} (for $w_0 w_a$ddm), and can be summarized as follows:
\begin{itemize}[noitemsep,topsep=0pt]
    \item When the R20 prior is used in combination with a compilation of the latest CMB measurements,
the $H_0$ tension is resolved at the $1\sigma$ level but the consideration of a dynamical dark energy component exacerbates the $S_8$ tension.  In these cases, there is agreement between Planck and R20, making the use of the latter safe. The solution of the $H_0$ tension is attributable to the phantom character of the dark energy.
\item  We do not find evidence for the disintegration of CDM into dark radiation in any of the different combinations of observational datasets under study,  i.e. $\alpha$ is always consistent with zero.
\end{itemize}
The corresponding graphical variations of free parameters in $w_0$ddm and $w_0 w_a$ddm models are displayed in Figs.~\ref{fig:wDDM} and \ref{fig:w0waDDM} respectively.  

In summary, we have shown that when the Planck 2018 temperature and polarization data are combined with other observations, such as the Hubble diagram of type Ia supernovae and the large-scale structure of the universe as traced by galaxies, there is no evidence for dark matter disintegrating at a
rate $Q \propto \mathcal{H} \rho_{\rm ddm}$. However, one may consider a more generalized version of the disintegration rate involving both $\rho_{\rm ddm}$ and $\rho_{\rm dr}$ and explore the impact of this new functional form on the detrmination of the cosmological parameters. We hope to report the results for different disintegration rates in the future.

\begin{acknowledgments}

LAA is supported by the U.S. National Science Foundation (NSF Grant PHY-2112527).
EDV acknowledges the support of the Addison-Wheeler Fellowship awarded by the Institute of Advanced Study at Durham University.
SP gratefully acknowledges the financial support given by the  Department of Science and Technology (DST), Govt. of India under the Scheme  
``Fund for Improvement of S\&T Infrastructure (FIST)'' [File No. SR/FST/MS-I/2019/41]. YW's work was supported by the National Natural Science Foundation of China under Grant No. 12075109, 11575075. 
WY's work was supported by the National Natural Science Foundation of China under Grants No. 11705079 and No. 11647153, and Liaoning Revitalization Talents Program under Grant no. XLYC1907098.
\end{acknowledgments}

\section*{Appendix}

To complete our direct comparison with the results of Ref.~\cite{Pandey:2019plg}, in this Appendix we combine Planck data with R20, despite the tension between them is above $3\sigma$ for the $\Lambda$ddm model. This combination should be considered therefore with caution, and this is the reason why it gives completely different results  and larger error bars with respect to the CMB alone case. Actually, the combination of CMB+R20 shifts the $H_0$ value reducing the tension to $1.4\sigma$  because of the prior assumed, while resolving the $S_8$ mismatch at the  $1\sigma$ level. This, however, is driven by the $H_0$ prior and manifests at the price of a $1\sigma$ indication for $\alpha\neq0$, i.e. $\alpha=0.0038^{+0.0019}_{-0.0023}$ at 68\% CL, and a much larger upper limits at 95\% CL than the other dataset combinations. This solution of the Hubble tension is in contrast with the intermediate-redshift data that reduce the effectiveness of this model to relax the $H_0$ and $S_8$ tensions. However, it is important to stress that when we compare our results (which include Planck polarization data at  high-$\ell$) with those in Ref.~\cite{Pandey:2019plg} (which do not include Planck polarization data at  high-$\ell$) we observed a measurable improvement on the constrained parameter space; namely, the mean value of $\alpha$ is shifted towards lower values and its uncertainty is reduced by about 30\%.

\bibliography{biblio}
\end{document}